\documentclass[nofootinbib,a4paper,aps,prd,10pt,superscriptaddress,showkeys,showpacs,twocolumn]{revtex4}

\usepackage{graphicx}
\usepackage{amsmath}
\usepackage{amsfonts}
\usepackage{amsthm}
\usepackage{mathrsfs}
\usepackage{amssymb}
\usepackage{epsfig}
\usepackage{amscd}
\usepackage{dcolumn}
\usepackage{bm}
\usepackage{natbib}
\usepackage{url}
\usepackage{xspace}

\usepackage[normalem]{ulem}
\usepackage{appendix}
\usepackage{algorithm}
\usepackage{algorithmic}

\usepackage{siunitx}
\usepackage{paralist}
\usepackage{placeins}

\usepackage{xcolor}

\usepackage{comment}

\usepackage{hyperref}
\hypersetup{%
    ,urlcolor=blue
    ,citecolor=blue
    ,linkcolor=blue
    }

\usepackage{etoolbox}
\makeatletter
\patchcmd{\frontmatter@RRAP@format}{(}{}{}{}
\patchcmd{\frontmatter@RRAP@format}{)}{}{}{}
\renewcommand\Dated@name{}
\makeatother

\def\be{\begin{equation}}
\def\ee{\end{equation}}
\def\bea{\begin{eqnarray}}
\def\eea{\end{eqnarray}}

\begin{document}

\title{Testing generalized logotropic models with cosmic growth}

\author{Kuantay~\surname{Boshkayev}}
\email[]{kuantay@mail.ru}
\affiliation{%
National Nanotechnology Laboratory of Open Type,\\
Al-Farabi Kazakh National University, Al-Farabi av. 71, 050040 Almaty, Kazakhstan.
}
\affiliation{%
Department of Physics, Nazarbayev University, 010000 Nur-Sultan, Kazakhstan.
}
\affiliation{%
Department of Engineering Physics, Satbayev University, 22 Satbayev str., 050013 Almaty, Kazakhstan.
}

\author{Talgar~\surname{Konysbayev}}
\email[]{konysbayev@aphi.kz}
\affiliation{%
National Nanotechnology Laboratory of Open Type,\\
Al-Farabi Kazakh National University, Al-Farabi av. 71, 050040 Almaty, Kazakhstan
}

\author{Orlando~\surname{Luongo}}
\email[]{orlando.luongo@unicam.it}
\affiliation{%
National Nanotechnology Laboratory of Open Type,\\
Al-Farabi Kazakh National University, Al-Farabi av. 71, 050040 Almaty, Kazakhstan
}
\affiliation{%
Dipartimento di Matematica, Universit\`a di Pisa, Largo B. Pontecorvo 5, 56127 Pisa, Italy
}
\affiliation{%
Divisione di Fisica, Universit\`a di Camerino, Via Madonna delle Carceri, 9, 62032 Camerino, Italy
}

\author{Marco~\surname{Muccino}}
\email[]{marco.muccino@lnf.infn.it}
\affiliation{%
National Nanotechnology Laboratory of Open Type,\\
Al-Farabi Kazakh National University, Al-Farabi av. 71, 050040 Almaty, Kazakhstan.
}
\affiliation{%
INFN, Laboratori Nazionali di Frascati, Via Enrico Fermi, 54, 00044 Frascati (RM), Italy
}

\author{Francesco~\surname{Pace}}
\email[]{francesco.pace9@unibo.it}
\affiliation{%
Dipartimento di Fisica ed Astronomia "Augusto Righi", Alma Mater
Studiorum Universit\`a di Bologna, Via Gobetti 93/2, I-40129 Bologna, Italy.
}
\affiliation{%
Jodrell Bank Centre for Astrophysics, School of Natural Sciences, University of Manchester, \\
Alan Turing Building, Oxford Road, Manchester M13 9PL, United Kingdom.
}

\date{(Received 18 March 2021; accepted 17 June 2021)}

\begin{abstract}
We check the dynamical and observational features of four typologies of logotropic dark energy models, leading to a thermodynamic cosmic speed up fueled by a single fluid that unifies dark energy and dark matter. We first present two principal Anton-Schmidt fluids where the Gr\"uneisen parameter $\gamma_{\rm G}$ is free to vary and then fixed to the special value $\gamma_{\rm G}=\tfrac{5}{6}$. We also investigate the pure logotropic model, corresponding to $\gamma_{\rm G}=-\frac{1}{6}$. Finally, we propose a new logotropic paradigm that works as a generalized logotropic fluid, in which we split the role of dark matter and baryons. We demonstrate that the logotropic paradigms may present drawbacks in perturbations, showing a negative adiabatic sound speed which make perturbations unstable. We thus underline which model is favored over the rest. The Anton-Schmidt model with $\gamma_{\rm G}=\frac{5}{6}$ is ruled out while the generalized logotropic fluid seems to be the most suitable one, albeit weakly disfavored than the $\Lambda$CDM model. To fix numerical constraints, we combine low- and higher-redshift domains through experimental fits based on Monte Carlo Markov Chain procedures, taking into account the most recent Pantheon supernovae Ia catalog, Hubble measurements and $\sigma_8$ data points based on the linear growth function for the large scale structures. We also consider two model selection criteria to infer the statistical significance of the four models under examination. We conclude there is a statistical advantage to handle the Anton-Schmidt fluid with the Gr\"uneisen parameter free to vary and/or fixed to $\gamma_{\rm G}=-\frac{1}{6}$. The generalized logotropic fluid indicates suitable results, statistically more favored than the other models until the sound speed is positive, becoming unstable in perturbations elsewhere. We emphasize that the $\Lambda$CDM paradigm works statistically better than any kind of logotropic and generalized logotropic models, while the Chevallier-Polarski-Linder parametrization is statistically comparable with logotropic scenarios. Finally, we propose that generalizing the Gr\"uneisen parameter by including the effects of temperature would guarantee the sound speed to be positive definite at all redshifts.
\end{abstract}



\maketitle

\section{Introduction}
The currently observed accelerated expansion of the Universe is widely supported by experimental evidence  \citep{1999ApJ...517..565P,1998Natur.391...51P,1998AJ....116.1009R,2003ApJ...594....1T,2003Sci...299.1532B,2003ApJS..148....1B,2003ApJS..148..135H,2003ApJS..148..161K,2003ApJS..148..175S,2005ApJ...633..560E}.
The concordance paradigm assumes that a fluid whose corresponding density, $\rho_{\rm de}$, under the form of a cosmological constant, $\Lambda$, with equation of state (EoS), say $\omega_{\rm de}=P_{\rm de}/\rho_{\rm de}\equiv\omega_{\Lambda}=-1$, sufficiently negative to counterbalance the action of gravity and to speed up the Universe today \citep{2000IJMPD...9..373S,2006IJMPD..15.1753C}. Any departure in terms of barotropic fluids stands for dark energy (DE) \citep{1998AJ....116.1009R,1999ApJ...517..565P,2003RvMP...75..559P,2013PhR...530...87W} with the purpose of overcoming the main caveats of the standard $\Lambda$CDM paradigm \citep{2003PhR...380..235P}, i.e., constructed using $\Lambda$. Explaining the DE nature passes through the use of first principles \citep{2012Ap&SS.342..155B,2003PhR...380..235P,2006IJMPD..15.2105S,2006IJMPD..15.1753C,  2000IJMPD...9..373S}, and/or in terms of extended/modified theories of gravity \citep{2017PhR...692....1N, 2007IJGMM..04..209S} and so on. All these approaches, although profoundly different, rely on the hypothesis that DE is an additional fluid different from baryons and dark matter (DM).\footnote{For a different perspective see, e.g., \cite{Luongo2018}.} Among the different possibilities of studying competing DE models \citep{2006IJMPD..15.1753C,2012Ap&SS.342..155B,2009IJMPA..24.1545M}, it could be possible to formulate some sort of thermodynamic acceleration, i.e., treating the Universe as a thermodynamic system, where thermodynamic considerations over the whole ensemble of fluids permit acceleration of the Universe today, adopting a single fluid that unifies DE and DM.

Relevant typologies of thermodynamic models that satisfy the above requirement are the so-called logotropic models, which have been introduced to overcome the cusp-core problem \citep{2015EPJP..130..130C}. These models attempt to unify the dark sector since the logotropic fluid recovers DE and/or DM in limiting regimes, in analogy to the Chaplygin gas \citep{Bertacca2007,Bertacca2008,Piattella2010,Bertacca2011}. A particular class of logotropic models has been recently introduced within the framework of Anton-Schmidt EoS \citep{AntonSchmidt1997, Mayeretal2003}. This class of models is similar to genuine logotropic paradigms \citep{2015EPJP..130..130C,2017PhLB..770..213F,2019EPJC...79..332B} and can be matched with modified versions of the Chaplygin gas. The advantage of this class of models is that the Anton-Schmidt EoS is physically interpreted as the deformation of the Universe under the action of cosmic expansion \citep{Capozziello} and describes the transition from a pressureless status (state) to one with negative pressure while satisfying the Debye approximation \citep{Debye1912}.

In this paper, we propose a proper treatment employing linear perturbations for a set of four classes of logotropic models, including Anton-Schmidt gases. Doing so, we analyze the dynamical and experimental features of two logotropic models and two Anton-Schmidt gases. In particular, we first consider the original version of the  logotropic fluid and then we introduce a new paradigm in which the logotropic counterpart explicitly distinguishes the role of DM from baryons. Afterwards, we investigate the most accredited versions of the Anton-Schmidt fluids, where the Gr\"uneisen parameter $\gamma_{\rm G}$ is constant throughout the Universe evolution. At first we allow it to be free and then we fix it to the special value, namely $\gamma_{\rm G}=\frac{5}{6}$, or alternatively $n=-\frac{1}{6}-\gamma_{\rm G}=-1$. We then demonstrate that Anton-Schmidt gases can be seen as generalized logotropic fluids. We, thus, investigate how structures evolve, generalizing the growth factor equation by taking into account the effects of the EoS and sound speed of each model.

To describe the evolution of the inhomogeneous energy shift we parameterize the growth function $f=\frac{\mathrm{d}\ln{\delta}}{\mathrm{d}\ln{a}}$ in terms of the growth index $\gamma$. Expanding in Taylor series, we get the corresponding approximate normalized growth function. Afterwards, numerical results are viewed in terms of Monte Carlo Markov Chain (MCMC) analyses based on the type Ia Supernova (SNe Ia) Pantheon data catalog, Hubble rates at different redshifts and redshift-space distortions, and through $\sigma_8$ data points based on the linear growth function for the large scale structures.

Inconsistencies among models are discussed with respect to the standard $\Lambda$CDM model. We show that the Anton-Schmidt gases are disfavored by intermediate redshift observations of the redshift space distortions with respect to pure logotropic models. Even though we demonstrate that logotropic models work better, we show statistical inconsistencies even for such scenarios with respect to both the $\Lambda$CDM, $\omega$CDM paradigms and Chevallier-Polarski-Linder (CPL) parametrization \cite{Chevallier2001,Linder2003}. 

We conclude that a possible solution to the above-raised issues of our underlying models could be to take a varying Gr\"uneisen index that depends upon the temperature, namely $\gamma_{\rm G}=\gamma_{\rm G}(T)$. This would enable the sound speed to be always positive definite throughout the Universe evolution, cancelling out any perturbation instabilities. 
The paper is thus structured as follows. In Sec.~\ref{sezione2} we exploit the concept of thermodynamic acceleration in the context of logotropic models and Anton-Schmidt gases. To do so, we highlight the basic properties of our four classes of thermodynamic models, confronting the genuine logotropic paradigm with Anton-Schmidt fluids. In Sec.~\ref{sezione3}, we work out linear perturbations for each model. We underline the basic differences and we evaluate the growth factor and the growth index $\gamma$. In Sec.~\ref{sezione4}, we present our fitting procedures, whose main results are analyzed and interpreted in Secs.~\ref{sezione4bis} and~\ref{sezione5}. Finally, in Sec.~\ref{sezione6}, we report our conclusions and perspectives.

\section{Thermodynamic dark energy: logotropic fluids}\label{sezione2}

Logotropic corrections to the Universe EoS are an attractive feature worth investigating. The original formulation of the Anton-Schmidt fluid can be clearly matched with logotropic DE models \citep{2015EPJP..130..130C} and Chaplygin gas \citep{2001PhLB..511..265K}. The simplest approach to determine a barotropic Anton-Schmidt EoS leads to \footnote{Please note that throughout the work we conventionally adopt natural units where $c=1$.}
\begin{equation}\label{eq:sample1}
 \omega(\rho) = A\left(\frac{\rho^{-n(T)-1}}{\rho_{\ast}^{-n(T)}}\right)\ln\left(\frac{\rho}{\rho_{\ast}}\right)\,,
\end{equation}
where $\omega$ is the background EoS $\omega \equiv P/\rho$, whereas $\rho_{\ast}$ and $\rho$ are the reference and matter densities, respectively. The constant $A$ is a normalization factor. The index $n$ depends on the absolute temperature $T$ of the environment, i.e., the Universe, and can easily be approximated to a constant in epochs where $T$ does not significantly evolve.

Arguably, from solid state physics, we can write $n =-{\frac{1}{6}}-\gamma_{\rm G}$, where $\gamma_{\rm G}$ is the Gr\"uneisen parameter, closely associated to the physical properties of the fluid itself (for details see, e.g., \citep{Gruneisen1912}). For $n = 0$, Eq.~(\ref{eq:sample1}) reduces to the genuine logotropic cosmological models \citep{2015EPJP..130..130C}. 

Considering the continuity equation in a Friedmann-Robertson-Walker  spacetime
\begin{equation*}
 \frac{\mathrm{d}\rho}{\mathrm{d}t}+3H(\rho+P) = 0\,,
\end{equation*}
one gets the Hubble parameter
\begin{equation}\label{eq:sample4}
 H^2 \equiv \left(\frac{\dot{a}}{a}\right)^2 = \frac{8 \pi G \epsilon}{3}\,,
\end{equation}
where one can split the total energy, $\epsilon$, into matter and DE counterparts \citep{Capozziello,Capozziello2}.

We distinguish four relevant cases. In the first, $n$ is a free parameter of the model, assumed constant throughout the cosmic evolution. This case corresponds to the simplest Anton-Schmidt gas, i.e., the one in which the Universe temperature has (small) influence over the evolution of the fluid itself. In the second case we set $n=-1$. This leads to a negligible effect of the temperature on $n$ that is no longer a free parameter of the model itself. Statistically speaking, \emph{a priori}, this case may be favored with respect to the first case and has been investigated in \citep{Capozziello2}. These two approaches correspond to limiting cases of the most general logotropic models, discussed as a third case. The latter represents a genuine logotropic paradigm, which simply requires $n\rightarrow0$. The last case is the formulation of a further logotropic model that we introduce, assuming a specific case of the pressure $P$, as we will detail below. For each model we will present below the most relevant expressions, needful to study the evolution of linear perturbations.

The total energy density can be split into two components, i.e., matter and DE as $\epsilon = \epsilon_{\rm m}+\epsilon_{\rm de}$ and the energy density parameters are defined as $\Omega_{\rm m,0}\equiv\epsilon_{\rm m,0}/\epsilon_{\rm c}$, $\Omega_{\rm de,0}\equiv\epsilon_{\rm de,0}/\epsilon_{\rm c} = 1-\Omega_{\rm m,0}$, where we used the critical energy density $\epsilon_{\rm c}\equiv 3H_0^2/(8\pi G)$.

Knowing the expression for the pressure $P$ and the density $\rho$, the adiabatic sound speed for a barotropic fluid is defined by
\begin{equation}\label{eq:cs2}
 c_{\rm s,a}^2 = \left(\frac{\partial P}{\partial a}\right)\left(\frac{\partial\rho}{\partial a}\right)^{-1}\,.
\end{equation}
This represents a key quantity, entering the linear perturbation equations, that determines the stability of perturbations.

Thus, to investigate the cosmological features of all our models, we employ the original formulation of the Anton-Schmidt pressure \cite{Capozziello2} and adopt the first law of thermodynamics, $\mathrm{d}\epsilon=\left(\frac{\epsilon+P}{\rho}\right)\mathrm{d}\rho$, which can be integrated as
\begin{equation}\label{lointegrale}
 \epsilon = \rho + \rho \int^{\rho}\mathrm{d}\rho^{\prime}\frac{P(\rho^{\prime})}{{\rho^{\prime}}^2}\,.
\end{equation}
Considering the pressure of a logotropic model, we can express $\epsilon$ in terms of $\rho$
\begin{equation}\label{eq:total rho}
 \epsilon = \rho + \frac{A}{2}\left(\frac{\rho}{\rho_{\ast}}\right)\ln^2{\left(\frac{\rho}{\rho_{\ast}}\right)}\,.
\end{equation}
Clearly, the limits for $a\ll 1$ and $a\gg 1$ lead to a matter and DE-dominated Universe, respectively.

Below we highlight the different models, concentrating on the total EoS, say $\omega=P/\epsilon$, the DE EoS, namely $\omega_{\rm de}=P_{\rm de}/\epsilon_{\rm de}$ and the adiabatic sound speed, as above defined by Eq.~(\ref{eq:cs2}).

\emph{Case I: ($n$ and $\gamma_{\rm G}$ as free coefficients)} For clarity, $n$ is a function of the temperature and so, in principle, it is free to vary throughout the cosmic evolution. However, given a particular cosmic era, it is plausible that $n$ only slightly evolves and, therefore, it could be considered roughly constant \cite{Capozziello}.

Assuming that deviations from the case of constant $n$ are negligible, we have
\begin{equation}\label{eq:sample72}
 \epsilon_{\rm de} = \epsilon_{\rm de,0}a^{3n} + \frac{3A}{n+1}\left(\frac{\rho_{\rm m,0}}{\rho_{\ast}}\right)^{-n}a^{3n}\ln{a}\,,
\end{equation}
where $\epsilon_{\rm m,0}$ and $\epsilon_{\rm de,0}$ are the matter and DE densities at current time, respectively. We further have 
\begin{equation}\label{eq:sampleT1}
 \epsilon_{\rm de,0} = -\frac{A}{n+1}\left(\frac{\rho_{\rm m,0}}{\rho_{\ast}}\right)^{-n} \left[\ln{\left(\frac{\rho_{\rm m,0}}{\rho_{\ast}}\right)}+ \frac{1}{n+1}\right]\,.
\end{equation}

\noindent Eq.~(\ref{eq:sample4}) can be written as
\be\label{eq:sample74}
 H^{(I)} = H_0\left[\Omega_{\rm m,0}a^{-3} + \Omega_{\rm de,0}\left(1+3B\ln{a}\right) a^{3n}\right]^{\frac{1}{2}}\,,
\ee
where the superscript $(I)$ denotes that the underlying quantity refers to model I. The characteristic parameter $B$ is 
\begin{equation}\label{eq:B}
 B \equiv -\left[\ln{\left(\frac{\rho_{\ast}}{\rho_{\rm m,0}}\right)} +\frac{1}{n+1}\right]^{-1}\,.
\end{equation}
The parameter $B$ is related to a dimensionless logotropic  temperature that is assumed to be constant for simplicity of computation. Further, we get
\bea
 \omega^{(I)} & = & -\frac{\Omega_{\rm de,0}\left[B+(n+1)\left(1+3B\ln{a}\right)\right] a^{3n}}{\Omega_{\rm m,0}a^{-3}+\Omega_{\rm de,0}\left(1+3B\ln{a}\right)a^{3n}}\,, \label{eq:sample79}\\
 \omega_{\rm de}^{(I)} & = & -\left(n+1\right)-\frac{B}{1+3B\ln{a}}\,, \label{eq:sample79b}\\
 c_{\rm s, a}^{2\,\,(I)} & = & \left(\frac{\Omega_{\rm de,0}}{\Omega_{\rm m,0}}\right)a^{3(n+1)} \times \nonumber \\
 && \left[(1+2n)B + n(n+1)(1+3B\ln{a})\right]\,,\label{eq:sample84}
\eea
respectively, the total EoS, the Anton-Schmidt EoS and the adiabatic sound speed. The latter is positive, leading to stable perturbations,  only if the argument of the second parenthesis is positive. This requirement is essential for structure formation theory.

\emph{Case II: (fixed $n=-1$ and $\gamma_{\rm G}=\tfrac{5}{6}$)} Here, the effect of the temperature does not influence the overall evolution. For this reason, $n$ is fixed to a precise value and the corresponding EoS aims at describing both the deceleration and acceleration epochs. 

Thus, handling the total density, splitting it into two contributions ($\epsilon = \epsilon_{\rm m} + \epsilon_{\rm de}$), with $\epsilon_{\rm m} = \epsilon_{\rm m,0}a^{-3}$, we have
\be\label{eq:rho_de}
 \epsilon_{\rm de} = \frac{\epsilon_{\rm de,0} }{a^3} - \frac{3A}{a^3}\left(\frac{\rho_{\rm m,0}}{\rho_{\ast}}\right)\ln{a}\ln{ \left(\frac{\rho_{\rm m,0}}{\rho_{\ast}}a^{-3/2}\right)}\,,
\ee
for the DE density. We defined
\begin{equation}
 \epsilon_{\rm de,0} = \frac{A}{2}\left(\frac{\rho_{\rm m,0}}{\rho_{\ast}}\right)\ln^2{\left(\frac{\rho_{m,0}}{\rho_\ast}\right)}\,.
\end{equation}

As for model I, we can define the characteristic parameter $B$
\be
 B\equiv\ln^{-1}\left(\frac{\rho_{\rm m,0}}{\rho_{\ast}}\right)\,,
\ee
which is quite different from the one given for the logotropic models
\be
 B_{\rm log}\equiv \left[\ln{\left(\frac{\rho_{\ast}}{\rho_{\rm m,0}}\right)}-1\right]^{-1}\,.
\ee

The latter, as stated above, depends on the logotropic temperature and holds a precise physical meaning. Contrary to its particular interpretation, in our present case, we expect to have $B<0$ as $\rho_{\ast}\gg 1$. Differently from previous results in the literature \cite{estuba}, we demonstrate here that the characteristic density is not necessarily the Planck density. Moreover, our experimental fits will show that, even considering the Planck density as extreme case for $\rho_{\ast}$, the model fails to predict high redshift evolution of the Universe.

Using the expression for $\epsilon_{\rm de,0}$, we can write the Hubble parameter as
\begin{equation}
 H^{(II)} = H_0\left[\Omega_{\rm m,0}a^{-3}+\Omega_{\rm de,0}(1-6B\ln{a}+9B^2\ln^2{a})a^{-3}\right]^{\frac{1}{2}}\,.
\end{equation}

As done for model I, we now present the expressions for the EoS of the total fluid and of the DE component and for the adiabatic sound speed, respectively:
\bea
 \omega^{(II)} & = & \frac{2B - 6B^2\ln{a}}{\Omega_{\rm de,0}^{-1}-6B\ln{a} + 9B^2\ln^2{a}}\,,\\
 \omega_{\rm de}^{(II)} & = & \frac{2B}{1-3B\ln{a}}\,,\\
 c_{\rm s, a}^{2\,\,(II)} & = & \frac{A\left[1+\ln{\left(\frac{\rho}{\rho_{\ast}}\right)}\right]}{\rho_{\ast}+\frac{A}{2}\left[2+\ln{\left(\frac{\rho}{\rho_{\ast}}\right)}\right]\ln{\left(\frac{\rho}{\rho_{\ast}}\right)}}\,.
\eea
For model II, we added the superscripts $(II)$, in analogy to model I. At small $z$, one gets $\omega_{\rm de}\approx 2B+6B^2(a-1)$, and the $\Lambda$CDM paradigm, for which $\omega_{de}\equiv-1$, is recovered when $B\rightarrow-1/2$ at $a=1$.

The expression for the sound speed can be written in terms of the characteristic parameter $B$. After simple manipulations we get
\begin{equation}\label{eq:cs nuovo}
 c_{\rm s, a}^{2\,\,(II)} = \frac{2B\Omega_{\rm de,0}(1+B-3B\ln{a})}{\Omega_{\rm m,0}}\,.
\end{equation}

Again, its sign depends upon the choice of the free constants here involved. This limitation of the model will reflect to our experimental fits.

\emph{Case III: (fixed $n=0$ and $\gamma_{\rm G}=-\frac{1}{6}$)} This case deals with a pure logotropic framework. In particular, we recover the third model as limit case of model I. Once again, the Universe temperature is here negligible with the peculiar choice $n=0$. This assumption unifies DE and DM as two byproducts of the same single fluid. 

As a possible physical justification of $n=0$, let us assume the DE EoS can be used for galactic dynamics. Hence, taking it within hydrodynamic equilibrium equations and noticing that DE and DM can be considered as single dark fluid, the pressure may describe both cosmic evolution and complicated galactic DM structures.

In particular, if DM halos are subject to hydrostatic equilibrium, in Newtonian regime we have
\begin{equation}\label{logojust1}
 \nabla P + \rho\nabla\Phi = {\bf 0}\,,
\end{equation}
and considering a polytropic relation $P=K\rho^{\gamma}$, one finds
\begin{equation}\label{logojust2}
 K\gamma \rho^{\gamma-1}\nabla \rho + \rho\nabla\Phi = {\bf 0}\,.
\end{equation}

To avoid a central cusp, the pressure should be constant. For the pressure gradient to counterbalance gravity, we require, however, $K\gamma>0$ and we can get a unified EoS by assuming $\gamma\rightarrow 0$ and $K\rightarrow \infty$. This leads to $A=K\gamma$, which is finite, and in such a limit we write
\begin{equation}\label{logojust3}
 A \frac{\nabla\rho}{\rho} + \rho\nabla\Phi = {\bf 0}\,.
\end{equation}
Comparing  Eqs.~(\ref{logojust1}) and~(\ref{logojust3}), it is evident that 
\begin{equation}\label{logojust4}
 P = A\ln{\rho} + C\,,
\end{equation}
where $A$ and $C$ are two integration constants. The logotropic EoS can be simplified fixing $C$, i.e., the cosmological constant contribution and so an easier form of the above equation for the pressure becomes
\be\label{eq:sampleT3}
 P = A\ln{\left(\frac{\rho}{\rho_{\ast}}\right)}\,,
\ee
where the cosmological constant has been suitably removed. 

The main purpose is now to determine the energy density $\epsilon$, knowing $\rho$. Assuming an adiabatic evolution, one immediately gets
\be\label{eq:sampleT0}
 \epsilon = \epsilon_{\rm m,0}a^{-3} + \epsilon_{\rm de,0} \left(1+3B\ln{a} \right)\,,
\ee
and the Hubble rate becomes
\be\label{eq:sampleT4}
 H^{(III)} = H_0 \left[\Omega_{\rm m,0}a^{-3} + \Omega_{\rm de,0}\left(1+3B\ln{a}\right)\right]^{\frac{1}{2}}\,,
\ee
that, clearly, can be recovered from  Eq.~(\ref{eq:sample74}) when $n=0$ and with the superscript $(III)$ that hereafter is used to distinguish model III from models I and II. Here, the parameter $B$ is defined as
\be\label{eq:sampleT5}
 B = \frac{A}{\epsilon_{\rm c}\Omega_{\rm de,0}}\,.
\ee

Analogously with the previous two paradigms, the relevant thermodynamic quantities for this model read
\begin{align}
 \omega^{(III)} = &\, -\frac{\Omega_{\rm de,0}\left(B+1+3B\ln{a}\right)} {\Omega_{\rm m,0}a^{-3} + \Omega_{\rm de,0}\left(1+3B\ln{a} \right)}\,, \label{eq:sampleT1a} \\
 \omega_{\rm de}^{(III)} = &\, -1-\frac{B}{1+3B\ln{a}}\,, \label{eq:sampleT1b}\\
 c_{\rm s, a}^{2\,\,(III)} = &\, B a^{3} \left(\frac{\Omega_{\rm de,0}}{\Omega_{\rm m,0}}\right)\,, \label{eq:sampleT2}
\end{align}
which represent the total EoS, that of the DE and the adiabatic sound speed, respectively. The $\Lambda$CDM model is recovered for $B=0$. Taking into account our previous calculations, we can easily relate the adiabatic sound speed in pure Anton-Schmidt's paradigm, model I, with Eq.~\eqref{eq:sampleT2} by
\begin{equation}\label{comparison}
 c_{\rm s,a}^{2\,\,(I)} = \frac{2(1+B)(3B\ln{a}-B-1)}{a^3}\, c_{\rm s,a}^{2\,\,(III)}\,,
\end{equation}
with $B$ given as in (\ref{eq:B}).

\emph{Case IV: (modified $n=0$ logotropic model)} The logotropic version of DE, considered in case III, takes into account the basic assumption that the DE EoS can be used for galactic dynamics.

However, the main disadvantage of models I, II and III is that their sound speeds are negative definite in broad regions of the Universe's evolution. This could be a direct consequence of how the model unifies DE with DM. A general solution of this issue, as we will stress below, is offered by removing the constant temperature approximation on $n$. This speculation deserves, however, accurate investigations since we do not know \emph{a priori} how the Gr\"uneisen index depends on the temperature. The prerogative of understanding which is the most suitable $\gamma_{\rm G}=\gamma_{\rm G}(T)$ function will be object of future investigations and would help to unify inflation with dark energy epochs.

Hence, we propose the simplest generalization of model III, where DM and baryons are unified with DE. This would guarantee the sound speed to be positive definite in the wider domain of the Universe's expansion unlike models I, II and III. Thus, assuming again the hydrostatic equilibrium, i.e., Eq.~\eqref{logojust1}, one can extend the polytropic EoS through a double polytropic of the form
\begin{equation}\label{doublelog}
 P = K_1\rho^{\gamma_1} + K_2\rho^{\gamma_2}\,,
\end{equation}
where only two constants, namely $(K_1,\gamma_1)$, behave as above, i.e., $K_1\gamma_1=const$, with $(K_1,\gamma_1)\rightarrow(\infty,0)$. The other two, namely $(K_2,\gamma_2)$, vary freely. In particular, we have 
\begin{equation}\label{logojustother}
 K_1\gamma_1 \rho^{\gamma_1-1}\nabla \rho + K_2\gamma_2 \rho^{\gamma_2-1}\nabla \rho+\rho\nabla\Phi = {\bf 0}\,.
\end{equation}

\noindent It is interesting to work out the case $(K_2,\gamma_2)\rightarrow(C, 1)$. In this respect, we get
\begin{equation}\label{extendedlogotropicmodel13}
 P = -A\ln{\left(\frac{\rho}{\rho_{\ast}}\right)} + C\left(\frac{\rho}{\rho_\ast}\right)\,,
\end{equation}
where $A\equiv -K_1\gamma_1$. For the sake of clarity, the above procedure can be extended up to an arbitrary order of polytropic equations of state. However, we limit our attention to the simplest case provided by the choice \eqref{doublelog} that has the intriguing advantage to reduce to $\sim \rho$ and $\sim\ln{\rho}$ for very large and small density $\rho$, respectively. Plugging Eq.~\eqref{extendedlogotropicmodel13} into \eqref{lointegrale}, we obtain
\begin{equation}\label{epvsrhocase4}
 \epsilon = \rho + \left[A + C\left(\frac{\rho}{\rho_{\ast}}\right)\right]\ln\left(\frac{\rho}{\rho_{\ast}}\right) + A\,.
\end{equation}

Note that $A$, although under the form of a cosmological constant, does not have the meaning of a cosmological constant contribution, since it is formally given as the product by $K_1$ and $\gamma_1$. Following the same strategy of the three previous models, we get $\epsilon \equiv \epsilon_{\rm b} + \epsilon_{\rm cdm} + \epsilon_{\rm de}$, explicitly showing the contributions of baryons ($\epsilon_{\rm b}$) and cold DM ($\epsilon_{\rm cdm}$) that arise from the term $\sim C\rho$, in Eq.~\eqref{extendedlogotropicmodel13}. We can then define the total matter contribution as $\rho_{\rm m} = \rho_{\rm cdm}+\rho_{\rm b}$. By comparison with case III, we impose  
\begin{subequations}
 \begin{align}
  \rho_{\rm b} & = \frac{\rho_{\rm b,0}}{a^3}\,,\\
  \rho_{\rm cdm} & = \frac{\rho_{\rm cdm,0}}{a^3} = \frac{C}{a^3}\left(\frac{\rho_{\rm b,0}}{\rho_\ast}\right)\ln\left(\frac{\rho_{\rm b,0}}{\rho_\ast}\right)\,,\\
  \rho & = A\left[1+\ln\left(\frac{\rho_{\rm b,0}}{\rho_\ast}\right)\right]-3\left[A+\frac{C}{a^3}\left(\frac{\rho_{\rm b,0}}{\rho_\ast}\right)\right]\ln{a}\,,
 \end{align}
\end{subequations}
where we defined
\begin{align}
 B = &\, -\left[1+\ln\left(\frac{\rho_{\rm b,0}}{\rho_{\ast}}\right)\right]^{-1}\,,\\ 
 A = &\, -\left(1-\Omega_{\rm m,0}\right)B \rho_{\rm c}\,.
\end{align}

\noindent We are now in the position to get the Hubble rate
\bea
 H^{(IV)} = H_0\Big[\Omega_{\rm m,0}a^{-3} & + & (1-\Omega_{\rm m,0})\left(1+3B\ln{a}\right) + \nonumber\\
 & + & \Omega_{\rm cdm,0}\left(\frac{3B}{B+1}\right)a^{-3}\ln{a}\Big]^{\frac{1}{2}}\,,
\eea
so that the total EoS, the DE EoS and the adiabatic sound speed are, respectively,
\bea
 \omega^{(IV)} & = & -1 - \frac{B\left(1-\Omega_{\rm m,0}\right)-\Omega_{\rm m,0}\,a^{-3}}{E^2} + \nonumber \\
 & & \qquad - \frac{\Omega_{\rm cdm,0}}{E^2}
 \left(\frac{B}{B+1}\right)a^{-3}\left(1-3\ln{a}\right)\,,\\
 \omega_{\rm de}^{(IV)} & = & -1 - \frac{B\left(1-\Omega_{\rm m,0}\right)}{E^2-\Omega_{\rm m,0}a^{-3}}+\nonumber\\
   & &-\frac{\Omega_{\rm cdm,0}}{a^3 E^2-\Omega_{\rm m,0}}\left(\frac{B}{B+1}\right)\left(1-3\ln{a} \right),\,\,\,\,\,\label{DEeos4}\\
 c_{\rm s,a}^{2\,\,(IV)} & = & \frac{B}{\Omega_{\rm b,0}} \left[\left(1-\Omega_{\rm m,0}\right)a^3 - \frac{\Omega_{\rm cdm,0}}{B+1}\right]\,. \label{soundspeedmod}
\eea
In the above expressions, the superscript $(IV)$ refers to model IV, $E=H/H_0$ and $\Omega_{\rm m,0}=\Omega_{\rm b,0}+\Omega_{\rm cdm,0}$.

\section{Evolution of the growth factor}\label{sezione3}

The models described in the previous section are all specific cases of a wider formalism where DM and DE are unified into a single dark fluid \citep{Bertacca2007,Bertacca2008,Piattella2010,Bertacca2011}. Both components represent limiting cases of the more general fluid at early and late times, respectively. This has two beneficial consequences: the first is that we only need a single component to explain both the structure formation and the observed accelerated expansion; the second is that we can treat DM and DE at the perturbation level in exactly the same way.

In this section, we concentrate on the analysis of linear perturbations, by deriving the appropriate differential equation for the growth factor. Note that, while formally perturbations are linear only at early times and/or on large scales, the growth factor equation is valid only on small scales, and a fundamental assumption is that it is valid on subhorizon scales. We also remind the reader that the growth factor is one of the main ingredients of the halo mass function, making it, therefore, an important quantity to study.

In the literature, the following expression is widely used
\begin{equation}
 \ddot{\delta} + 2H\dot{\delta} - 4\pi G \rho_{\rm m} \delta = 0\,,
\end{equation}
where $\rho_{\rm m}$ is the (total) matter density. This equation, valid on small scales and for linear perturbations, implicitly assumes that matter is the clustering component. In this case, $c_{\rm s}^2=\omega=0$. However, our setup is more general than that and the correct way of proceeding needs to take into account the additional degrees of freedom of our model. In this respect, our physical setup is very similar to what has been recently done for generalized DM \citep{Kopp2016,Thomas2016} by \cite{Pace2020}. One of the differences with respect to that work is that in our case both the background EoS and the sound speed are, in general, time-dependent and both need to be properly evaluated. We have no freedom to set one of them to zero to simplify our expressions.

Before starting, we need to define correctly which perturbations are we talking about. From our previous discussion, we said that our fluid can be decomposed into two components, one representing matter and the other one resembling a smooth DE component driving the accelerated expansion of the Universe. Of these two fluids, only DM is clustering and, therefore, modifications to the standard growth factor equation will be related to the DM component.

To derive our equations, we start from the equations for a fluid with pressure and density perturbations, so that $c_{\rm s}^2=\delta P/\delta\rho$ and follow the derivation in \cite{Pace2017}. However, since the models considered here are adiabatic, the sound speed entering into perturbations is the adiabatic one, i.e., $c_{\rm s}^2=c_{\rm s,a}^2$. This will enormously simplify our analysis, in that we do not need to use any scalar field descriptions for the models.

The continuity and Euler equations read, respectively,
\begin{align}
 \delta^{\prime} + 3(s-w)\delta + (1+w)\tilde{\theta} = &\, 0\,, \label{eq:continuity}\\
 \tilde{\theta}^{\prime} + \left(2+\frac{H^{\prime}}{H}\right)\tilde{\theta} + \frac{3}{2}(1+3s)\Omega(a)\delta = &\, 0\,, \label{eq:euler}
\end{align}
where the prime represents the derivative with respect to $\ln{a}$, $\delta=\delta\rho/\rho$ is the dimensionless density perturbation of the fluid, $s=c_{\rm s,a}^2$ is the dimensionless adiabatic sound speed for perturbations and $\tilde{\theta}=\theta/H$, where $\theta$ represents the divergence of the peculiar velocity $\boldsymbol{u}$, $\theta=\vec{\nabla}\cdot\boldsymbol{u}$. In the Euler equation, $\Omega(a)$ represents the energy density parameter of the perturbed fluid, i.e., the matter component, so that $\Omega(a) = \Omega_{\rm m}(a)$, which is defined as
\begin{equation}
 \Omega_{\rm m}(a) = \frac{\rho_{\rm m}}{\rho_{\rm m}+\rho_{\rm de}}\,.
\end{equation}

We are now in a position to derive a second order equation for $\delta$. To do so, we take the derivative of Eq.~(\ref{eq:continuity}) and we substitute in it Eq.~(\ref{eq:euler}). For simplicity of notation, we define the following variables:
\begin{equation*}
 A = 3(s-w)\,, \quad B = 1+\omega\,, \quad f = \frac{3}{2}(1+3s)\Omega(a)\,.
\end{equation*}

\noindent The final equation then reads
\begin{equation}\label{eq:gf}
 \delta^{\prime\prime} + \left(A\delta\right)^{\prime} + 
 \left[\left(2+\frac{H^{\prime}}{H}\right)-\frac{B^{\prime}}{B}\right]
 (\delta^{\prime}+A\delta) - Bf\delta = 0\,.
\end{equation}
This expression is similar in its form, and fully equivalent, to what obtained in \cite{Abramo2007}.

We can now specify the two free functions in Eq.~\eqref{eq:gf}. Since the clustering component is DM, we set $\omega=0$, whereas the adiabatic sound speed is applied to the whole model. Note the specularity with clustering DE models, where $c_{\rm s}^2=0$ to allow a clustering similar to that of DM. We can then simplify Eq.~(\ref{eq:gf}) to
\begin{equation}\label{eq:gf1}
 \delta^{\prime\prime} + 3\left(s\delta\right)^{\prime} + 
 \left(2+\frac{H^{\prime}}{H}\right)(\delta^{\prime} + 3s\delta) - \frac{3}{2}(1+3s)\Omega_{\rm m}(a)\delta = 0\,.
\end{equation}

It is often interesting to consider the logarithmic derivative of the growth factor $f=\mathrm{d}\ln{\delta}/\mathrm{d}\ln{a}$. Its equation, in light of the modified Eq.~(\ref{eq:gf1}), reads
\begin{equation}\label{eqn:exact_f}
 f^{\prime} + f^2 + 3(s^{\prime}+sf) + \left(2+\frac{H^{\prime}}{H}\right)(f+3s) - \frac{3}{2}(1+3s)\Omega_{\rm m} = 0\,.
\end{equation}

For many models, it is possible to give a phenomenological solution for $f$ \citep{Paul}
\begin{equation}\label{eq:f}
 f \approx \Omega_{\rm m}^{\gamma}(a)\,,
\end{equation}
where $\gamma$ is the so-called growth index.

To study the evolution of $\gamma$, the simplest thing to do is to plug the approximate solution for $f$ into (\ref{eq:f}). This leads to a first-order differential equation for $\gamma$ which reads:
\begin{align}
\label{totalgamma}
 \gamma^{\prime} + & \frac{3\Omega_{\rm de}\omega_{\rm de}}{\ln{\Omega_{\rm m}}}\gamma + \frac{\Omega_{\rm m}^{\gamma}}{\ln{\Omega_{\rm m}}} + 3\frac{s+s^{\prime}\Omega_{\rm m}^{-\gamma}}{\ln{\Omega_{\rm m}}} \\
 + & \frac{1-3\Omega_{\rm de}\omega_{\rm de}}{2\ln{\Omega_{\rm m}}}\left(1+3s\Omega_{\rm m}^{-\gamma}\right) - \frac{3(1+3s)}{2\ln{\Omega_{\rm m}}}\Omega_{\rm m}^{1-\gamma} = 0\,. \nonumber
\end{align}

To make progress in solving this equation, we will make some assumptions which allow us to linearize it. When $\Omega_{\rm m}\approx\mathcal{O}(1)$, we can write $\ln{\Omega_{\rm m}}\approx-\Omega_{\rm de}$ and $\Omega_{\rm m}^{\gamma}\approx 1-\gamma\Omega_{\rm de}$. Under these approximations, the evolution of $\gamma$ is described by
\begin{align}\label{eq:gamma_linear}
\gamma^{\prime} + & \left[1-3\omega_{\rm de}+\frac{3}{2}+\frac{3}{2}s\left(2+3\Omega_{\rm de}\omega_{\rm de}\right)-3s^{\prime}\right]\gamma = \nonumber \\
& \frac{3}{2}(1+6s)(1-\omega_{\rm de}) + \frac{3s^{\prime}}{\Omega_{\rm de}}\,,
\end{align}
and we will assume that this approximation is roughly valid also at later times.

For $s=0$, assuming constant $\omega_{\rm de}$, $\gamma$ recovers the solution of the $\omega$CDM model
\begin{equation}\label{eq:gamma_wCDM}
 \gamma = \frac{3(\omega_{\rm de}-1)}{6\omega_{\rm de}-5}\,,
\end{equation}
which reduces to $\gamma=6/11$ for the standard $\Lambda$CDM paradigm \citep{1998ApJ...508..483W}.

The equation above has a formal solution which can be expressed via the integral of the coefficient of $\gamma$ and of the source term. However, this expression will provide a very limited insight into the physics of the model.

Whether an analytical solution is possible or not, Eq.~(\ref{eq:gamma_linear}) is showing that the adiabatic sound speed acts as a correction to the standard picture. To better see why this is the case, we will also assume that both $\omega_{\rm de}$ and $s$ are constant. We will further consider, consistently for the derivation of the equation for $\gamma$, that $2+3\Omega_{\rm de}\omega_{\rm de}\approx 2$. Under these assumptions, Eq.~(\ref{eq:gamma_linear}) becomes
\begin{equation}
\gamma^{\prime} + \left(1-3\omega_{\rm de}+\frac{3}{2}+3s\right)\gamma = \frac{3}{2}(1+6s)(1-\omega_{\rm de})\,,
\end{equation}
whose solution is
\begin{equation}\label{eq:gamma_s_w}
 \gamma = \frac{3(1+6s)(1-\omega_{\rm de})}{5+6(s-\omega_{\rm de})}\,.
\end{equation}

The evolution of $\gamma$ for our models will be discussed in the next sections, in comparison with the predictions of the standard model, i.e. $\gamma=\frac{6}{11}$, of the $\omega$CDM, through the use of Eq.~\eqref{eq:gamma_s_w}, and evolving DE. Finally, we stress that Eq.~\eqref{eq:gamma_s_w} provides a prominent role in approximating $\gamma$ for models I, II and III as we will outline below. 

\section{Experimental limits}\label{sezione4}

One of the main purposes of this paper is to understand which model better approximates the Universe dynamics among the four paradigms described above. 

Understanding which model is effectively the most suitable one to pass through higher redshift data domains in which the degeneracy problem is somehow healed is needed. In particular, to fix cosmological bounds over the different paradigms, we employed the standard low-redshift data surveys based on: observational Hubble data set (OHD) \citep{2002ApJ...573...37J}, SNe Ia with the {\it Pantheon} catalog \citep{2018ApJ...859..101S} and higher-redshift points coming from the data based on the use of the so-called growth function $f$ for large scale structure, together with the normalization of matter power spectrum,  $\sigma_8$.

\subsection{Likelihood analysis}

Here, we perform a set of MCMC analyses involving all the above cases. The best set of parameters is hereafter dubbed {\bf x}, entering the total log-likelihood function, $\ln{\mathcal{L}}$
\begin{equation}
 \ln{\mathcal{L}} = \ln{\mathcal{L}_{\rm OHD}} + \ln{\mathcal{L}_{\rm SN}} + \ln{\mathcal{L}_{\rm f}} + \ln{\mathcal{L}_{\rm \sigma_8}}\,.
\end{equation}

Below, we introduce the log-likelihood for each of the probes.

\begin{table}[!t]
 \begin{center}
  \renewcommand*{\arraystretch}{0.8}
  \caption{\label{tab:HOD} $H(z)$ measurements for the OHD data used in the text, reported in units km\,s$^{-1}$\,Mpc$^{-1}$, with uncertainties (second column). In the first column we report the observed redshift $z$, whereas in the third column we report the reference paper in which the corresponding measurement has been first presented.}
  \begin{ruledtabular}
   \begin{tabular}{|ccc|}
    $z$ &$ H \pm \sigma_H$ &  Reference \\
    \hline
    0.0708  & $69.0  \pm 19.68$ & \cite{Zhang2014} \\
    0.09    & $69.0  \pm 12.0$  & \cite{Jimenez2002} \\
    0.12    & $68.6  \pm 26.2$  & \cite{Zhang2014} \\
    0.17    & $83.0  \pm 8.0$   & \cite{Simon2005} \\
    0.179   & $75.0  \pm 4.0$   & \cite{Moresco2012} \\
    0.199   & $75.0  \pm 5.0$   & \cite{Moresco2012} \\
    0.20    & $72.9  \pm 29.6$  & \cite{Zhang2014} \\
    0.27    & $77.0  \pm 14.0$  & \cite{Simon2005} \\
    0.28    & $88.8  \pm 36.6$  & \cite{Zhang2014} \\
    0.35    & $82.1  \pm 4.85$  & \cite{Chuang2012} \\
    0.352   & $83.0  \pm 14.0$  & \cite{Moresco2016} \\
    0.3802  & $83.0  \pm 13.5$  & \cite{Moresco2016} \\
    0.4     & $95.0  \pm 17.0$  & \cite{Simon2005} \\
    0.4004  & $77.0  \pm 10.2$  & \cite{Moresco2016} \\
    0.4247  & $87.1  \pm 11.2$  & \cite{Moresco2016} \\
    0.4497  & $92.8  \pm 12.9$  & \cite{Moresco2016} \\
    0.4783  & $80.9  \pm 9.0$   & \cite{Moresco2016} \\
    0.48    & $97.0  \pm 62.0$  & \cite{Stern2010} \\
    0.593   & $104.0 \pm 13.0$  & \cite{Moresco2012} \\
    0.68    & $92.0  \pm 8.0$   & \cite{Moresco2012} \\
    0.781   & $105.0 \pm 12.0$  & \cite{Moresco2012} \\
    0.875   & $125.0 \pm 17.0$  & \cite{Moresco2012} \\
    0.88    & $90.0  \pm 40.0$  & \cite{Stern2010} \\
    0.9     & $117.0 \pm 23.0$  & \cite{Simon2005} \\
    1.037   & $154.0 \pm 20.0$  & \cite{Moresco2012} \\
    1.3     & $168.0 \pm 17.0$  & \cite{Simon2005} \\
    1.363   & $160.0 \pm 33.6$  & \cite{Moresco2015} \\
    1.43    & $177.0 \pm 18.0$  & \cite{Simon2005} \\
    1.53    & $140.0 \pm 14.0$  & \cite{Simon2005} \\
    1.75    & $202.0 \pm 40.0$  & \cite{Simon2005} \\
    1.965   & $186.5 \pm 50.4$  & \cite{Moresco2015} \\
   \end{tabular}
  \end{ruledtabular}
 \end{center}
\end{table}

\begin{itemize}

\item[(a)]\emph{Hubble rate likelihood}: To evaluate the Hubble rate likelihood, we notice that OHD points are cosmology-independent measurements of the Hubble rate at various $z$ through the differential age method \cite{2002ApJ...573...37J,ohd1}.
The Hubble rate is written by the identity $H(z)=-(1+z)^{-1}\Delta z/\Delta t$. Thus, from spectroscopic measurements of the age difference $\Delta t$, and redshift difference $\Delta z$, of couples of passively evolving galaxies that formed at the same time one infers the set of Hubble points \cite{ohd2}. The corresponding log-likelihood function is then given by
\bea
 \ln{\mathcal{L}_{\rm OHD}} & = & -\frac{1}{2}\sum_{i=1}^{N_{\rm OHD}}\ln{\left(2\pi\sigma_{\rm H_i}^2\right)} \nonumber\\
 && -\frac{1}{2}\sum_{i=1}^{N_{\rm OHD}}\left[\frac{H_{\rm i}-H\left({\bf x},z_i\right)}{\sigma_{\rm H_i}}\right]^2\,, \label{chisquared1}
\eea
where $N_{\rm OHD}$ corresponds to the OHD data points, as reported in Table~\ref{tab:HOD}.

\item[(b)] \emph{Pantheon likelihood}: The Pantheon data set is the most updated SN Ia sample composed of $1048$ sources \citep{2018ApJ...859..101S}. The standardization of their light curves involves the following corrections: 1) the luminosity-stretch coefficient $\alpha$ and factor $\mathcal{X}_1$ and 2) the luminosity-colour coefficient $\beta$ and factor $\mathcal{C}$, and the distance corrections $\Delta_{\rm M}$ and $\Delta_{\rm B}$, based on SN host galaxy mass and predicted biases, respectively. Once these corrections are applied, all SN Ia light curves become standard and the associated distance moduli are defined as
\begin{equation}
 \mu = m_{\rm B} - \left(\mathcal{M} -\alpha \mathcal{X}_1 + \beta \mathcal{C} - \Delta_{\rm M} - \Delta_{\rm B}\right)\,,
\end{equation}
where $\mathcal{M}$ is the $B$-band absolute magnitude and $m_{\rm B}$ the $B$-band apparent magnitude \citep{2011ApJS..192....1C}. The log-likelihood function is given by \citep{2001A&A...380....6G}
\begin{equation}\label{eqn:chimarg}
 \ln{\mathcal{L}_{\rm SN}} = -\frac{1}{2}\left(a + \ln{\frac{e}{2 \pi}} - \frac{b^2}{e}\right)\,,
\end{equation}
where $a\equiv\Delta\vec{\mathbf{\mu} }^{T}\mathbf{C}^{-1}\Delta\vec{\mathbf{\mu} }$, $b\equiv\Delta\vec{\mathbf{\mu} }^{T}\mathbf{C}^{-1}\vec{\mathbf{1}}$, and $e \equiv
\vec{\mathbf{1}}^T\mathbf{C}^{-1} \vec{\mathbf{1}}$, in which $\Delta\mu\equiv \mu - \mu_{\rm th}\left({\bf x},z\right)$ is the vector of residuals between the observed distance moduli, $\mu$, and the theoretical ones, $\mu_{\rm th}$. Finally, $\mathbf{C}$ is the covariance matrix which is related to statistical and intrinsic systematic uncertainties of SNe \citep{2011ApJS..192....1C}. For the sake of clarity, with SNe Ia only the $H_0$ value cannot be constrained. This is another reason to combine such a data set with the others, presented here.

\item[(c)] \emph{Matter growth likelihood}: In linear theory, the growth of matter fluctuations is described by the growth function $f$, defined by Eqs.~(\ref{eqn:exact_f}) and (\ref{eq:f}). The log-likelihood is therefore given by \cite{mat1,mat2}
\be\label{flikelihood}
 \ln{\mathcal{L}_{\rm f}} = -\frac{1}{2}\sum_{i=1}^{N_{\rm f}}\left\{\ln{\left(2\pi\sigma_{\rm f_i}^2\right)} + \left[\frac{f_{\rm i}-f\left({\bf x},z_i\right)}{\sigma_{\rm f_i}}\right]^2\right\}\,.
\ee

\item[(c)] \emph{$\sigma_8$ likelihood}: An alternative observational probe of $\delta(z)$ is the rms mass fluctuation $\sigma_8(z)$. It is linked to $\delta(z)$ via
\begin{equation}
 \sigma_8(z) = \sigma_8(0)\frac{\delta(z)}{\delta(0)} = \sigma_8(0) e^{\int^{\frac{1}{1+z}}_{1}\Omega_{\bf m}^\gamma(a)\mathrm{d}a}\,,
\end{equation}
where $\sigma_8(0)$ is the value at $z=0$. Most of the currently available data points originate from the observed redshift evolution of the flux power spectrum of the Ly--$\alpha$ forest \cite{mat1,mat2}. To avoid the use of the additional parameter, $\sigma_8(0)$, in the fitting procedure, an alternative parameter can be used, i.e.,
\begin{equation}
 s_8(z_1,z_2) = \frac{e^{\int^{\frac{1}{1+z_1}}_{1}\Omega_{\bf m}^\gamma(a)\mathrm{d}a}}{e^{\int^{\frac{1}{1+z_2}}_{1}\Omega_{\bf m}^\gamma(a)\mathrm{d}a}}\,.
\end{equation}
The corresponding log-likelihood becomes
\bea
 \ln{\mathcal{L}_{\rm s8}} & = & -\frac{1}{2}\sum_{i=1}^{N_{\rm s_8}}\ln{\left(2\pi\sigma_{\rm s_{8,i}}^2\right)} \nonumber\\
 && -\frac{1}{2}\sum_{i=1}^{N_{\rm s_8}}\left[\frac{s_{\rm 8,i}-s_8\left({\bf x},z_i,z_{i+1}\right)}{\sigma_{\rm s_{8,i}}}\right]^2\,,\label{chisquared2}
\eea
where $\sigma_{\rm s_{8,i}}$ is derived by error propagation from the errors of $s_8(z_i)$ and $s_8(z_i+1)$.
\end{itemize}

\begin{table}[!t]
 \caption{Statistical comparison among the $\Lambda$CDM, $\omega$CDM paradigms, the CPL parametrization, and models I--IV. The reference scenario is $\Lambda$CDM model, so that $\Delta{\rm AIC(BIC)}={\rm AIC(BIC)}_{\rm i}-{\rm AIC(BIC)}_{\rm \Lambda CDM}$.}
 \centering
  \begin{tabular}{|l|ccccc|}
   \hline
   \hline
   Model & $\ln \mathcal{L}_{max}$ & AIC & BIC & $\Delta$AIC & $\Delta$BIC\\
   \hline
   \hline
   $\Lambda$CDM & $-640.24$ & $1284.48$ & $1294.50$ & $0$ & $0$\\
   $\omega$CDM & $-640.48$ & $1286.97$ & $1301.99$ & $2.48$ & $7.49$\\
   CPL & $-640.22$ & $1288.44$ & $1308.48$ & $3.96$ & $13.98$\\
   \hline
   I & $-639.35$ & $1284.69$ & $1299.72$ & $0.21$ & $5.22$\\
   II & $-895.91$ & $1795.83$ & $1805.84$ & $511.32$ & $511.32$\\
   III & $-640.30$ & $1284.64$ & $1294.66$ & $0.16$ & $0.16$\\
   IV & $-640.29$ & $1284.58$ & $1294.59$ & $0.09$ & $0.09$\\
   \hline
   \hline
  \end{tabular}
 \label{tab:params1}
\end{table}

\section{Statistical analyses}\label{sezione4bis}

We now present the statistical performances of the various models through the use of the Bayesian selection criteria to measure the evidence of a given model against a reference scenario \citep{Kunz2006}, conventionally chosen as the $\Lambda$CDM paradigm. Specifically, we consider the Akaike information criterion (AIC) \citep{Akaike1974} and the Bayesian information criterion (BIC) \citep{Schwarz1978}, defined, respectively, as
\begin{subequations}
 \begin{align}
 \text{AIC} \equiv & -2\ln{\mathcal{L}}_{\rm max} + 2p \,, \\
 \text{BIC} \equiv & -2\ln{\mathcal{L}}_{\rm max} + p\ln{N}\,.
\end{align}
\end{subequations}
Here, $\mathcal{L}_{\rm max}$ is the maximum likelihood estimate, $p$ is the number of free parameters of the model, and $N$ the total number of data points. We note that, for high $N$, the BIC criterion penalizes the model with a large number of free parameters more severely than AIC.

Using these definitions, we calculated the differences $\Delta$AIC and $\Delta$BIC with respect to the reference scenario to measure the amount of information lost by adding extra parameters in the statistical fitting. Negative values of $\Delta$AIC and $\Delta$BIC would indicate that the model under investigation performs better than the reference model, while for positive values, one needs to know that

\begin{itemize}\label{AICintervals}
  \item[(a)] $\Delta\text{AIC(BIC)}\in [0,2]$ indicates a weak evidence in favour of the reference model, leaving open the question on which model is the most suitable one;
  \item[(b)] $\Delta\text{AIC(BIC)}\in (2,6]$ indicates a mild evidence against the given model with respect to the reference paradigm;
  \item[(c)] $\Delta\text{AIC(BIC)}> 6$ indicates a strong evidence against the given model, which should be rejected.
\end{itemize}

We report the $\Delta$AIC and $\Delta$BIC values for the different cosmological models in Table \ref{tab:params1}. They assume positive values for all the models indicating that data are either weakly or strongly in favor of the $\Lambda$CDM model. In more detail, according to the AIC criterion, the DE models (CPL and $\omega$CDM) are quite disfavored with respect to the reference model, while model II is strongly disfavored. Models I, III and IV are only weakly disfavored, in that $\Delta$AIC$<1$, so one can not really establish which model is favored. According to the BIC criterion, all the models are rather strongly disfavored (once again, model II in particular), but III and IV are slightly favored for the $\Lambda$CDM model, as $\Delta$BIC$<1$.

\begin{table*}[!t]
 \caption{Best-fit parameters for flat $\Lambda$CDM, $\omega$CDM, CPL, and models I--IV with $1$--$\sigma$ (and $2$--$\sigma$) errors. Hereafter, $h_0\equiv H_0/100s^{-1}$Mpc/km.}
 \centering
  \begin{tabular}{|l|cccccc|}
   \hline
   \hline
   Model & $h_0$ & $\Omega_{\rm m,0}$ & $\Omega_{\rm cdm,0}$ & $n$ & $w_0$& $w_1$\\
   \hline
   $\Lambda$CDM & $0.695^{+0.027 (+0.045)}_{-0.027 (-0.043)}$ & $0.290^{+0.028 (+0.047)}_{-0.026 (-0.042)}$ & - & - & - & - \\
   $\omega$CDM & $0.693^{+0.033 (+0.047)}_{-0.033 (-0.049)}$ & $0.270^{+0.047 (+0.070)}_{-0.040 (-0.056)}$ & - & - & $-0.92^{+0.10 (+0.15)}_{-0.13 (-0.21)}$ & - \\
   CPL & $0.690^{+0.039 (+0.057)}_{-0.036 (-0.049)}$ & $0.269^{+0.052 (+0.076)}_{-0.051 (-0.071)}$ & - & - & $-1.02^{+0.27 (+0.37)}_{-0.19 (-0.22)}$ & $0.63^{+0.72 (+0.74)}_{-1.79 (-2.54)}$\\
   \hline
   I & $0.689^{+0.034 (+0.053)}_{-0.033 (-0.046)}$ & $0.308^{+0.040 (+0.062)}_{-0.039 (-0.057)}$ & - & $0.004^{+0.015 (+0.027)}_{-0.009 (-0.022)}$ & - & - \\
   II & $0.492^{+0.017 (+0.028)}_{-0.017 (-0.028)}$ & $0.510^{+0.070 (+0.118)}_{-0.067 (-0.109)}$ & - & -1 & - & - \\
   III & $0.695^{+0.027 (+0.044)}_{-0.028 (-0.044)}$ & $0.291^{+0.028 (+0.046)}_{-0.026 (-0.042)}$ & - & 0 & - & - \\
   IV & $0.693^{+0.028 (+0.045)}_{-0.028 (-0.045)}$ & $0.291^{+0.029 (+0.049)}_{-0.028 (-0.045)}$ & $0.269^{+0.029 (+0.049)}_{-0.028 (-0.045)}$ & - & - & - \\
   \hline
   \hline
  \end{tabular}
 \label{tab:params2}
\end{table*}

\begin{figure*}[!ht]
 \centering
 \includegraphics[width=1.2\columnwidth,clip]{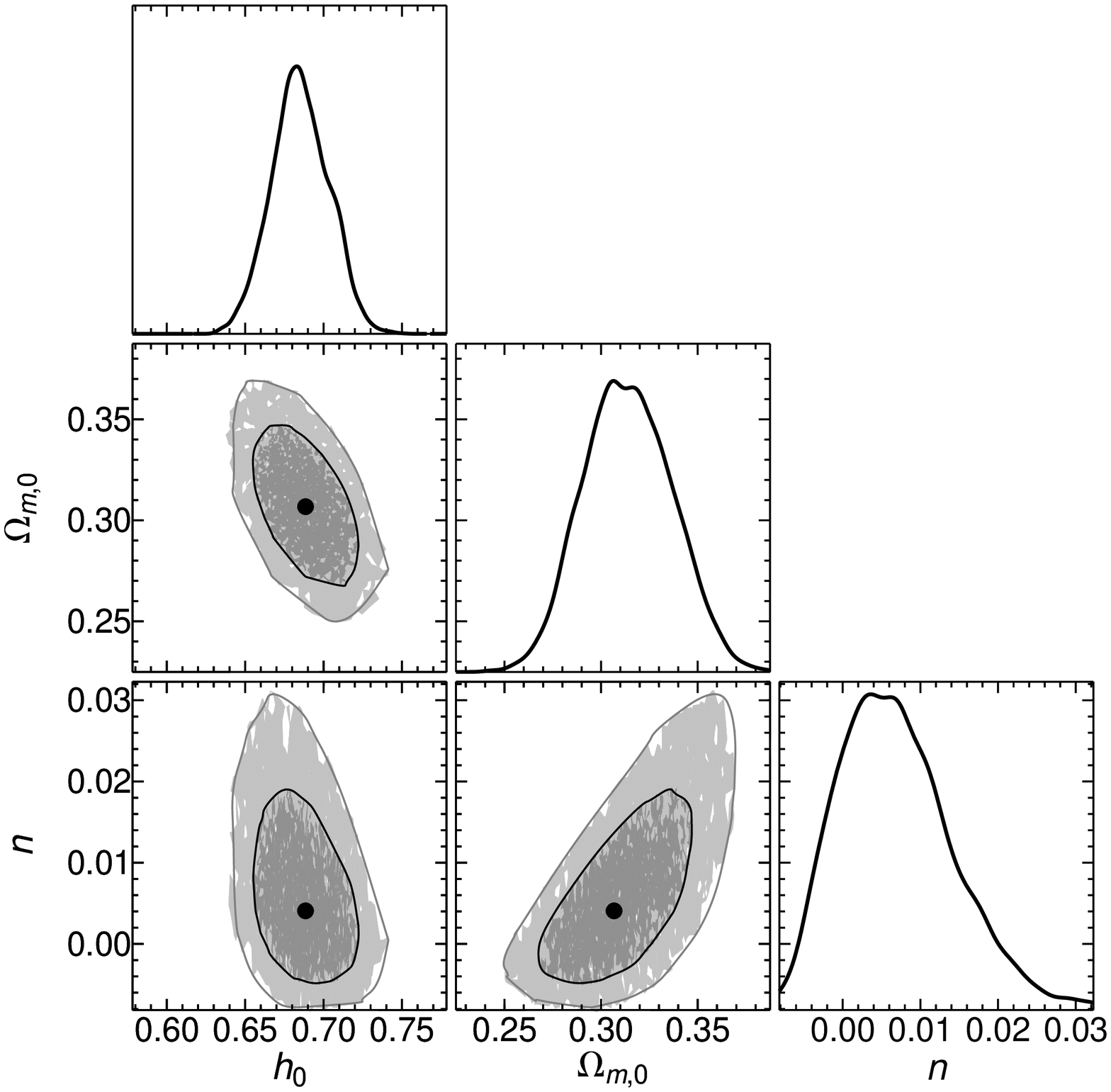}\hfill
 \includegraphics[width=0.8\columnwidth,clip]{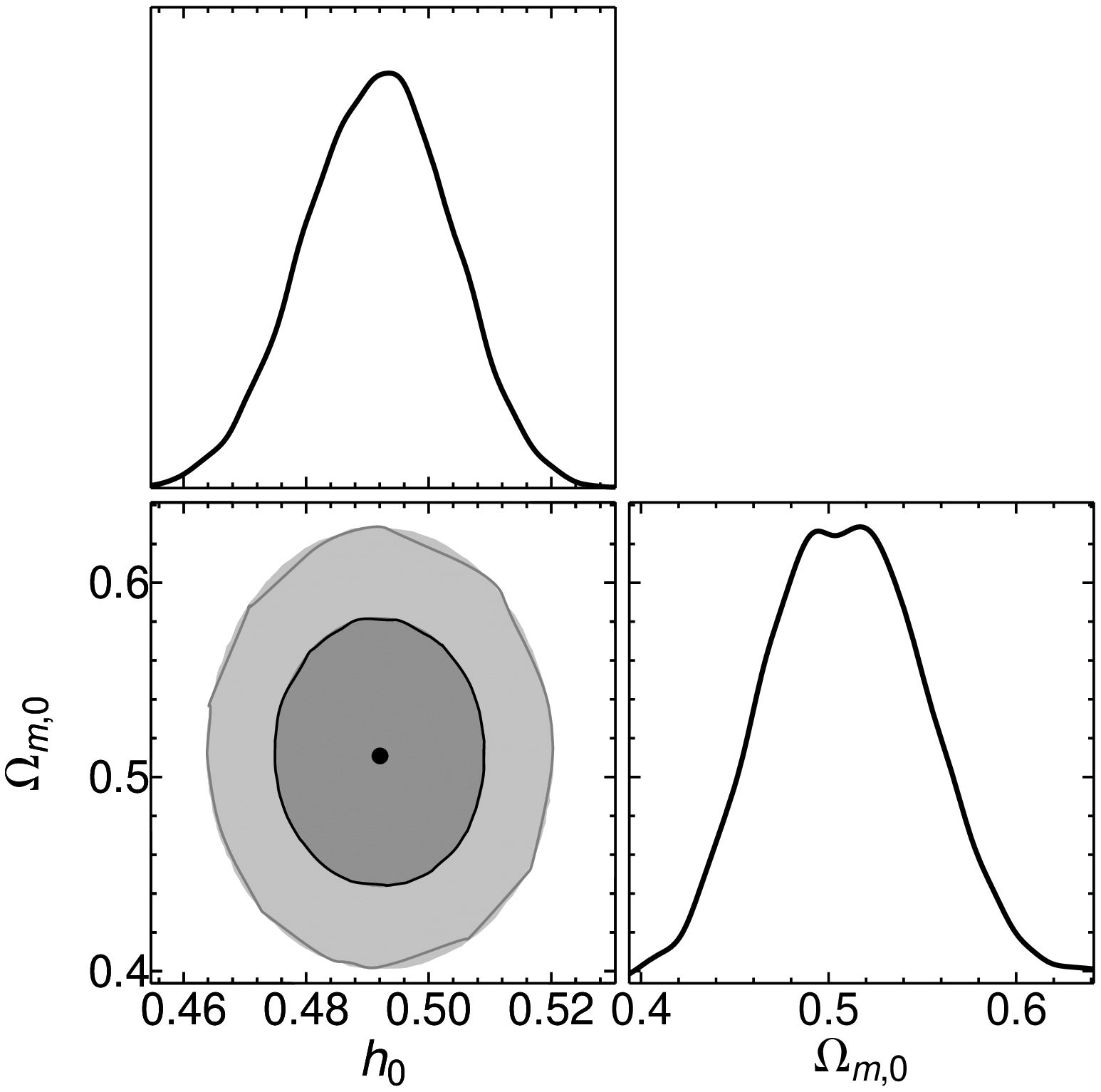}
 \includegraphics[width=0.8\columnwidth,clip]{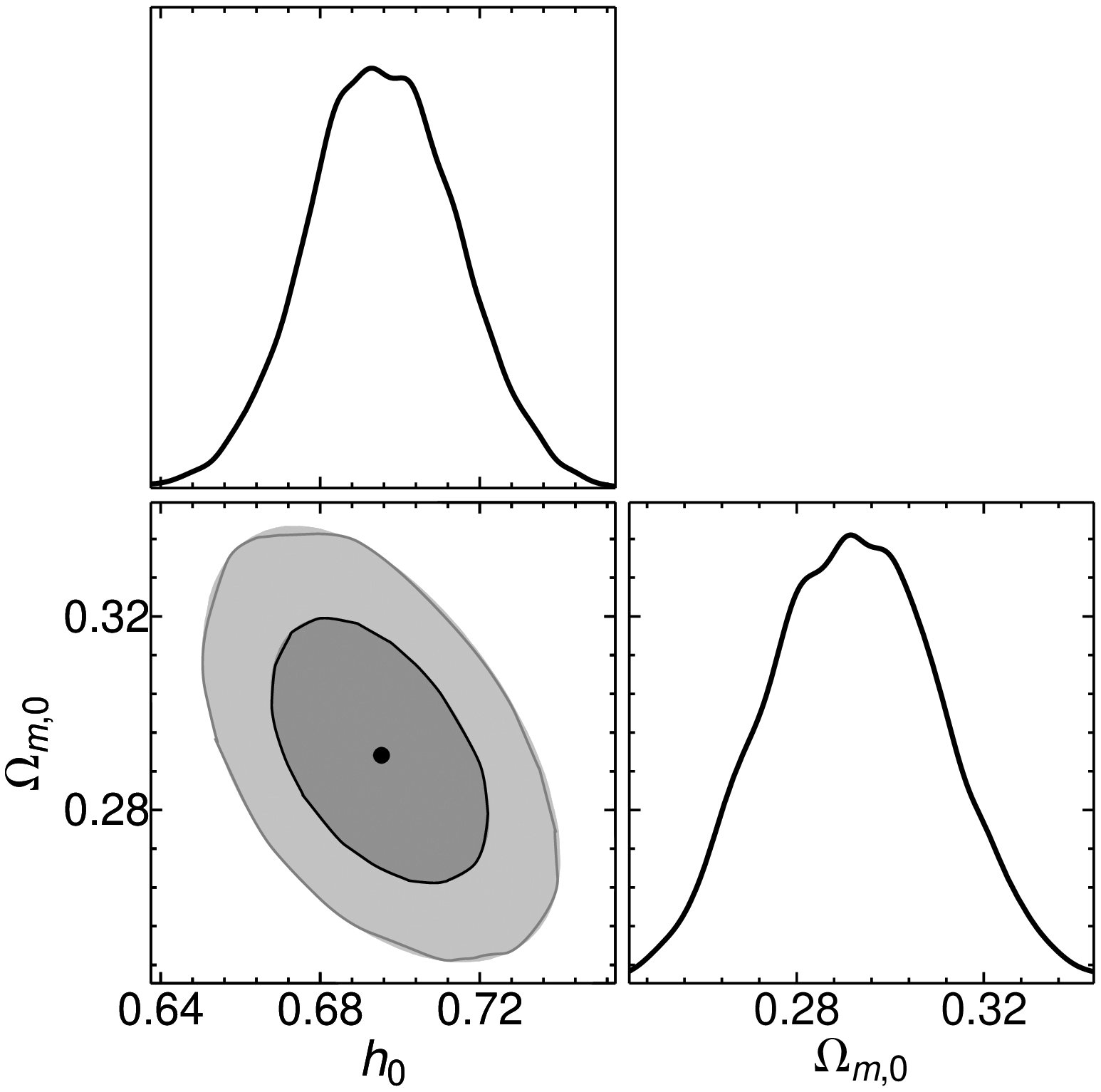}\hfill
 \includegraphics[width=0.8\columnwidth,clip]{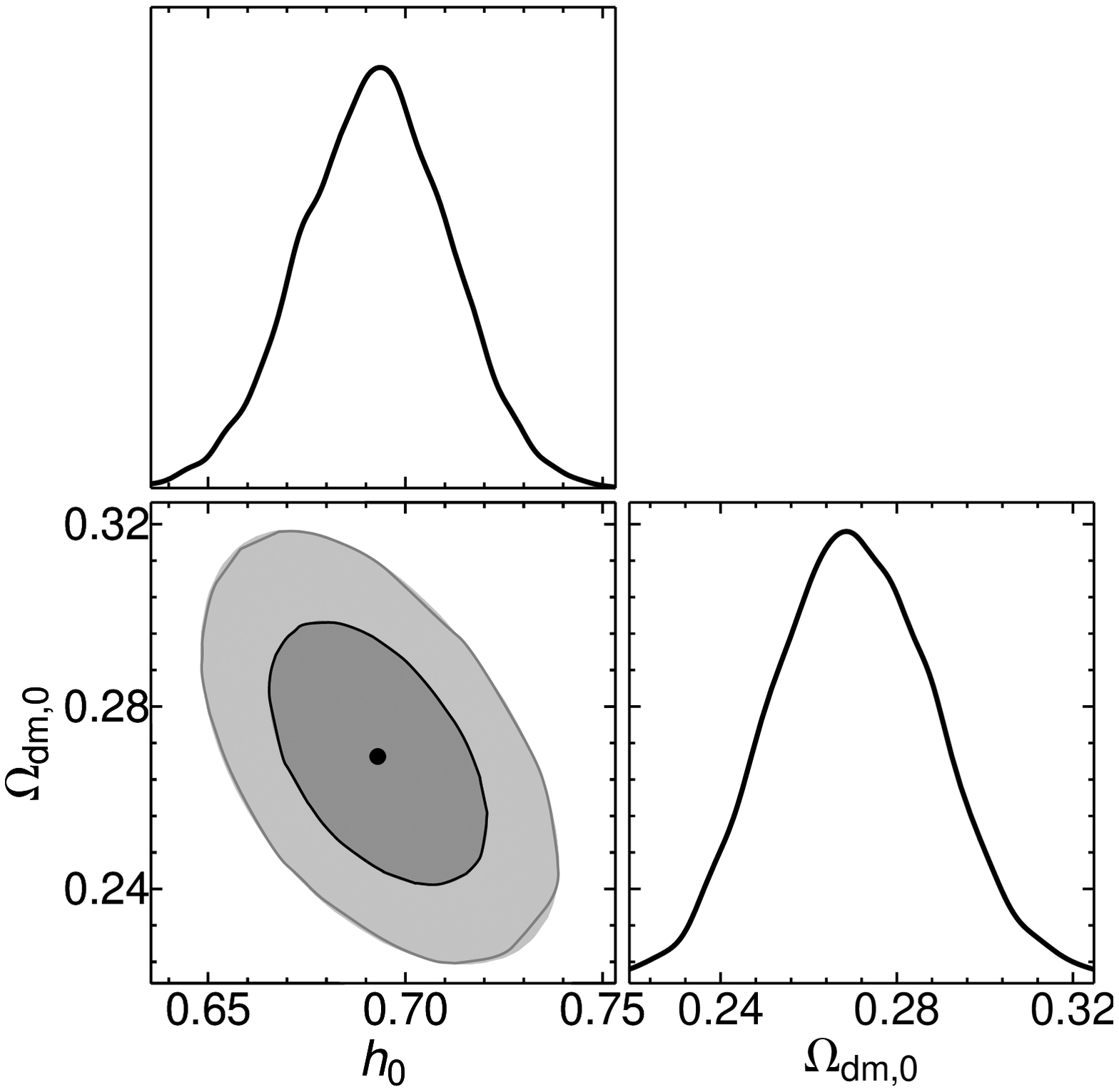}
 \caption{$1$--$\sigma$ (dark gray) and $2$--$\sigma$ (light gray) contour plots and best-fit parameters (black dot). In the top left (right) panel we show model I (II), while in the bottom left (right) panel, we present model III (IV).}
 \label{fig:MCMC}
\end{figure*}

\section{Discussion on theoretical results}\label{sezione5}

In this section, we describe all our numerical findings. They have been reported in Table~\ref{tab:params2}, where we included best fit parameters, and in Table~\ref{tab:params1}, for the AIC and BIC values. We also presented the results of our MCMC runs in Fig.~{\ref{fig:MCMC}}, where we show models I and II in the top panels and models III and IV in the bottom panels. 

In all our fits, $\rho_{\star}$ corresponds to the Planck density,  $\rho_{\rm Pl}$, in agreement with theoretical predictions. This assumption works well even for high redshift data, except for model II. In fact, the choice $n=-1$ is clearly ruled out by using the intermediate redshift data that we employed. 

Moreover, for model II even if $\rho_{\star}\neq\rho_{\rm Pl}$ (letting $\rho_{\star}$ be free to vary) the model does not work well with our data catalogs at small and intermediate redshifts.

Hubble constant constraints, for all our models, except again for model II, appear in tension with what found by \cite{Riess2019}. The error bars up to $2$--$\sigma$ confidence levels are not large enough to avoid the tension with Riess measurements at $\geq 1$--$\sigma$. In principle such a result may be affected by systematics and may be reconsidered in view of future developments. In all our cases, the Hubble parameters are compatible with Planck results \cite{Planck2016}.

Focusing on each model, we can compare our results with the concordance $\Lambda$CDM paradigm, and with two evolving DE scenarios, i.e., here the $\omega$CDM model and the CPL parametrization. In model I, $n$ severely differs from the outcomes obtained in \cite{Capozziello}. This is closely related to the higher redshift data sets that we adopted here. This fact, if confirmed by future analyses, suggests the first model is non-unequivocally constrained by cosmic data, leading to a clear limitation of the model itself. In particular, the value of $n$ is positive but very small. We show that if one adopts only small redshift data, the model is quite unbounded and the predictions over $n$ are unconstrained, i.e., severely different from our findings. In addition, here $n$ is compatible with zero at both $1$-- and $2$--$\sigma$ confidence levels. The values for both matter density $\Omega_{\rm m,0}$ and Hubble parameter are compatible with the $\Lambda$CDM scenario at $1-\sigma$ confidence level, making it, according to the AIC criterion, only slightly disfavored. 

Our numerical results remark that negative $n$ are excluded for Anton-Schmidt frameworks, in disagreement with previous studies, e.g., \cite{Capozziello}. In view of this, it is easy to stress again that even statistical criteria establish that model II is highly disfavored by data.

The genuine logotropic paradigm, model III, works quite well in describing the Universe dynamics. Both matter density and the Hubble function are in good agreement with the values found for the reference $\Lambda$CDM model. As a consequence, both the AIC and the BIC are small, $(\sim 0.16)$, showing that the $\Lambda$CDM model is only weakly favored. The genuine logotropic model is very similar to model IV, in which one includes an additional parameter, the baryon density $\Omega_{\rm b}$. As it is not possible to constrain $\Omega_{\rm b}$ within our redshift range, we take it fixed to the Planck constraint, $\Omega_{\rm b}\simeq 0.02242$. The pure logotropic model is largely more predictive than Anton-Schmidt paradigms, certifying that: 
\begin{itemize}
 \item[(a)] in general,  Anton-Schmidt paradigms (models I and II) seem to be less predictive than pure logotropic models (models III and IV),
 \item[(b)] generalized versions of the pure logotropic model (model III) do not  significantly change the experimental expectations over the free coefficients. 
\end{itemize}
An immediate interpretation of the above two points is the following: the Anton-Schmidt models are disfavored than pure logotropic scenarios because of the approximations made on $n$. In fact, as already stated above, in the original Anton-Schmidt picture, $n$ is a general function of the temperature. This fact could severely influence the goodness of Anton-Schmidt models, for example enabling the sound speed to be always positive definite.

However, comparing the AIC and BIC criteria as reported in Table~\ref{tab:params1}, we notice only a weak evidence for the statistical goodness of logotropic models with respect to Anton-Schmidt paradigms. The statistical significance of logotropic models, for both models III and IV, is lower than the concordance paradigm, as a consequence of the additional free parameters. In other words, the standard cosmological model is still favored in describing the cosmic expansion history even with respect to genuine logotropic models. Foe the sake of completeness, it is interesting to note that for models I (top left panel), III and IV (bottom panels) there is a small anticorrelation between $h_0$ and $\Omega_{\rm m}$, while this does not occur for model II. This is one of the reasons why this model performs so badly. Note that there is also a positive correlation between the free index $n$ and the matter density parameter $\Omega_{\rm m,0}$: this means that it is difficult to modify $n$ without finding unreasonable values of $\Omega_{\rm m,0}$.

Analogous comparisons can be performed among our models and evolving DE frameworks. So, regarding the CPL parametrization, both parameters ($w_0$ and $w_1$) are compatible with the cosmological constant scenario, i.e., $w_0=-1$ and $w_a=0$. Whilst the value of $h_0$ is very similar to what we found for our reference model, the matter density parameter is about $7\%$ smaller, which corresponds to less than $1$--$\sigma$ difference. Nevertheless, this model has two additional parameters with respect to the $\Lambda$CDM model and is, therefore, heavily penalized by the selection criteria. This explains the large $\Delta$AIC and $\Delta$BIC values, compared with the concordance paradigm. Similar considerations can be done for the $\omega$CDM model, for which we found values very similar to the CPL parametrization. It is interesting to notice that with $w_0=-0.92$, the model is in the ``quintessence regime", but still compatible to a $\Lambda$CDM model within $1$--$\sigma$. So, concerning the $\omega$CDM model, the $\Delta$AIC and $\Delta$BIC are lower than for the CPL one, making it statistically preferred with respect to the latter. This is clearly due to the smaller number of parameters. Comparing our paradigms with these results show that models I, III and IV seem to be more statistically favored than CPL and $\omega$CDM frameworks. However, we note that the statistical significance is weak and there is no reason \emph{a priori} to imagine that these models behave better than $\omega$CDM and CPL throughout the Universe evolution.

\subsection{Properties of the best fit models}
We now study the background properties, the adiabatic sound speed for linear perturbations and the evolution of the growth factor and index for our models.

\begin{figure}[!ht]
\centering
\includegraphics[width=\columnwidth,clip]{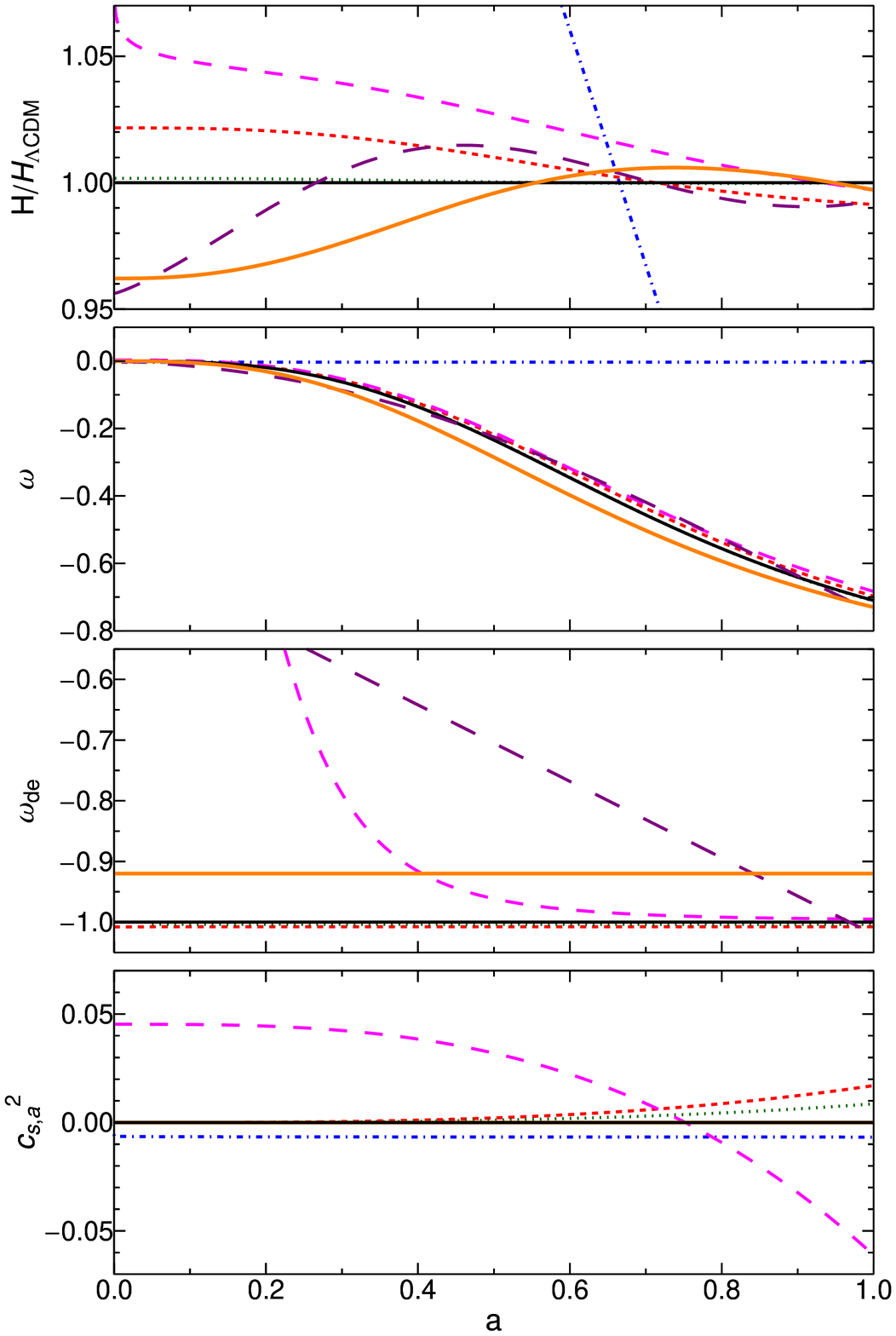}
\caption{Plots of the thermodynamic functions described for our four models in terms of $a$. Hereafter, for brevity we removed any superscripts. From top to bottom: the Hubble rate $H(a)$, normalized with the $\Lambda$CDM model, $H_{\Lambda{\rm CDM}}$, the total EoS parameter, $\omega$, the DE EoS, $\omega_{\rm de}$, and the adiabatic sound speed, $c_{\rm s,a}^2$. The values of the chosen parameters  are summarized in Table~\ref{tab:params2}.}
\label{ris:image01}
\end{figure}

\begin{figure}[!ht]
\centering
\includegraphics[width=\columnwidth,clip]{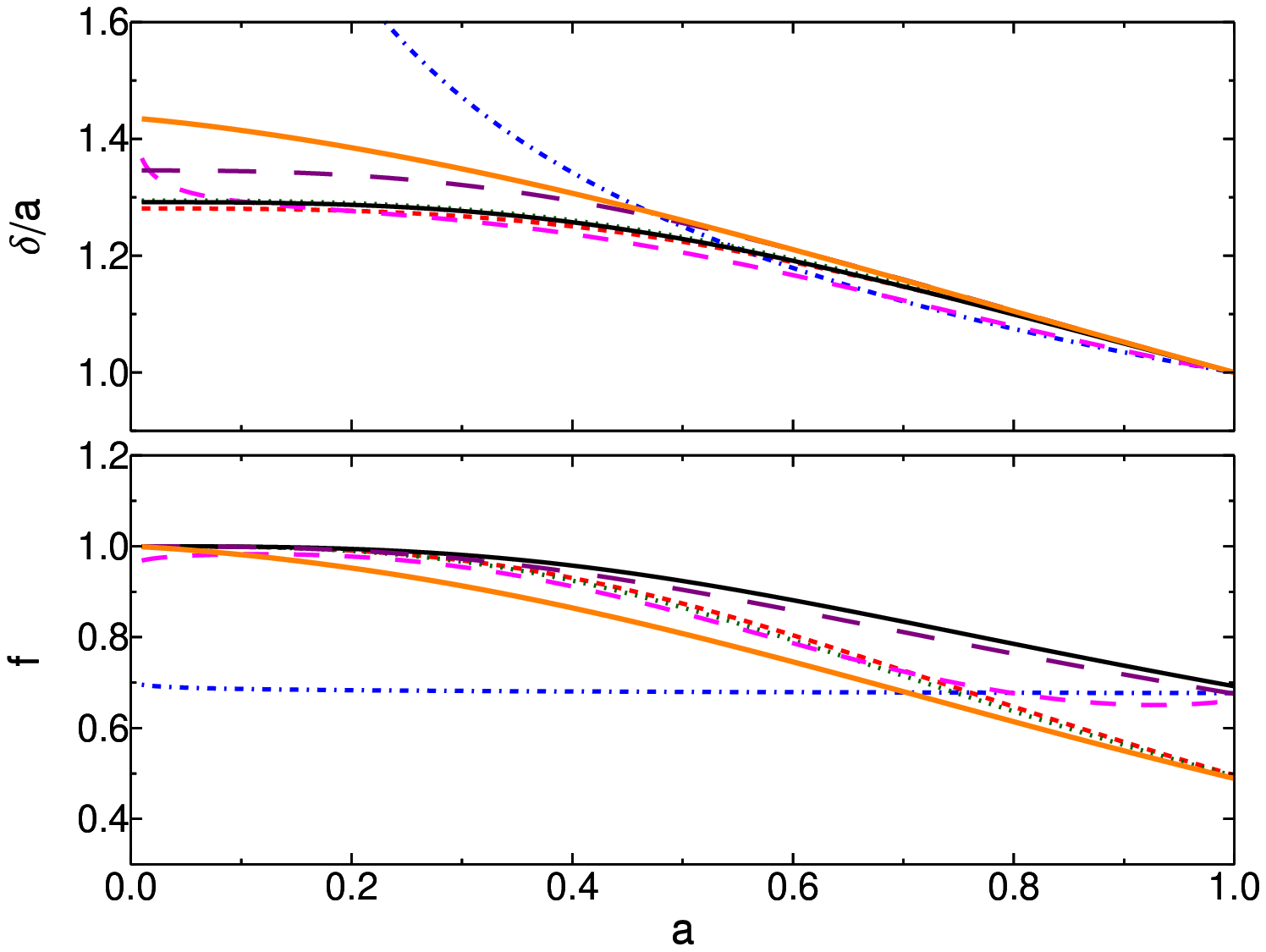}
\caption{The linear density perturbations over the scale factor $\delta/a$ (top panel) and the growth $f$ (bottom panel) as a function of $a$ for the four models. The values of the parameters chosen to plot each model are summarized in Table~\ref{tab:params2}.}
\label{ris:image02}
\end{figure}

In Fig.~\ref{ris:image01}, we present the evolution of the Hubble parameter normalized with the reference $\Lambda$CDM model (upper panel), the total EoS $\omega$ (second panel), the EoS for the DE component only, namely $\omega_{\rm de}$ (third panel), and the adiabatic sound speed (bottom panel) for the four models underlined in this work. For comparison, we also plot the expectations obtained from the $\Lambda$CDM, $\omega$CDM and CPL scenarios. From a qualitative point of view, model II with $n=-1$ is the one showing the largest differences with respect to the other logotropic models and to the two DE models considered in this work. Analyzing the background evolution from a more quantitative point of view, we note that model II deviates from the others at all redshifts, with differences much higher than 10\%. All the other models differ by at most a few percent, with a maximum of 5\% at early times. It is worth noticing that at late time the CPL model expansion rate is higher than for the $\Lambda$CDM model, while for $a<0.2$ it is a few percent smaller.

In the third panel from top, we present the evolution of the DE EoS $\omega_{\rm de}$. As expected, for the $\Lambda$CDM ($\omega$CDM) model, $\omega_{\rm de}=-1$ ($\omega_{\rm de}=-0.92$) constant throughout the cosmic history and for the CPL model $\omega_{\rm de}$ grows steadily as $w_1=0.63$. The evolution of the logotropic models is instead more interesting. For models I--III, the EoS of the DE component is rather constant and very close to $-1$, therefore having a behavior similar to that of the cosmological constant, $\Lambda$. Model IV, instead, is close to the cosmological constant up to $a=0.6$ and at earlier times it grows rapidly such that at $a=0.2$, we have $\omega_{\rm de}=-0.5$. This discussion helps us to understand the behavior of the total EoS of our models, presented in the second panel from top. In particular, for model II, the total EoS is practically null over the whole cosmic history, explaining why the model represents a very poor fit to the data.

The quantity of more interest for the evolution of perturbations is the adiabatic sound speed of each models. For adiabatic models, this quantity determines the stability of a given model against perturbations and ultimately its validity. 
This happens because, in general, we can write the evolution of perturbations as
\begin{equation}
 \ddot{\delta} + A(t)\dot{\delta} + c_{\rm s}^2\delta = 0\,,
\end{equation}
where $A(t)$ represents a generic damping term associated with the cosmic expansion. We also generically denoted the sound speed as $c_{\rm s}^2$, as this discussion has general validity. When $c_{\rm s}^2>0$, the solution is a damped harmonic oscillator, hence, perturbations are stable and bound. On the contrary, when $c_{\rm s}^2<0$, its solution is an exponential function and perturbations grow unbounded. Thus, the model is unstable and we can see the developing instability in the top panel. The instability is evident at very early times, where the quantity $\delta/a$ appears to be diverging. This is also reflected in the growth where $f$ is approximately constant. In fact, if $f\approx c_f$, with $c_f$ constant, then $\delta\propto a^{c_f}$ at all times (at odds with the physical intuition that an accelerated expansion of the Universe would lead structures to grow less).

All the models, except IV, have a roughly constant sound speed, which allows us to determine an approximate solution for the growth index $\gamma$, as shown in Sec.~\ref{sezione3}. Models I and III have a small (close to zero) sound speed. At early time the sound speed is  small and clustering properties are due to DM, while at late time the sound speed slightly grows, to reach the value of $\approx 2\times 10^{-2}$. Model II has a similar behavior, with an adiabatic sound speed squared constant but negative ($c_{\rm s}^2\approx -10^{-2}$). This implies that the model is unstable and explains once more why the model performs badly against the observational data. Finally, for model IV, at early times, the adiabatic sound speed is positive and of the order of $5\times 10^{-2}$, but at late times it becomes negative, reaching in absolute value approximately the same value at early times. Since the sound speed is in general also small when positive, the effects of instabilities at late times can be substantial.

\begin{figure}[!t]
\centering
\includegraphics[width=\columnwidth,clip]{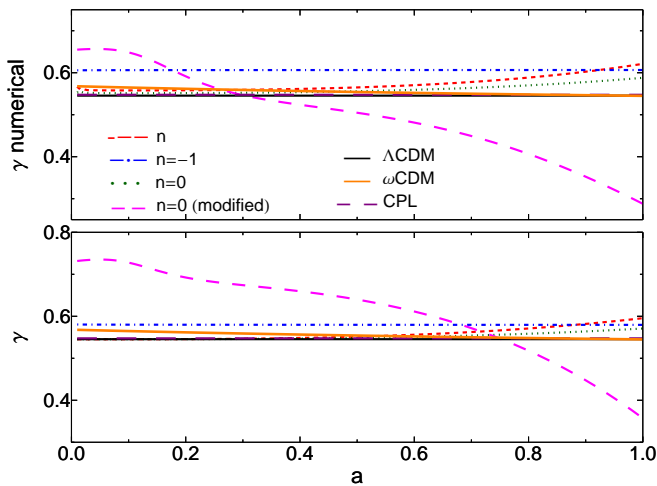}
\caption{The growth index $\gamma$ computed numerically from Eq.~\eqref{totalgamma} (top panel) and from the analytical solution in Eq.~\eqref{eq:gamma_s_w} (bottom panel) as a function of $a$ for the four models. The parameters chosen to plot each model are summarized in Table~\ref{tab:params2}. Each model is displayed, as indicated in the legend. Line styles will be the same for all the following figures.}
\label{ris:image03}
\end{figure}

The discussion about the sound speed is a useful introduction to the study of the evolution of the density perturbations and the growth. Models I and III, as expected from the previous analysis (especially in terms of Bayesian evidence) show a behavior very close to that of the $\Lambda$CDM model in terms of $\delta$, but differ more in the evolution of $f$ (see Fig.~\ref{ris:image02}). This is easily understood remembering that the growth is directly proportional to the time derivative of $\delta$ and that differential quantities highlight differences. It is, therefore, crucial, to also consider this quantity when comparing models. Note that model II is fully nonpredictive as stressed several times in this work and $f_{\rm CPL}\simeq f_{\rm \Lambda CDM}$. We now consider model IV. To do so we notice that throughout the cosmic history, perturbations grow similarly to the $\Lambda$CDM model. It is, however, more interesting to consider its logarithmic derivative, $f$. Even if its evolution is qualitatively different from that of the reference model, we notice a peculiar behavior at very early and late times. Thus, at early times, $f$ first flattens and is of the order unity as expected, but then it tends to decrease. This is symptomatic of a departure from a limiting Einstein-de Sitter regime in that the sound speed at early times is small but non-negligible. At late times, we see the effects of a negative sound speed: since perturbations now grow unbounded, the late time cosmic acceleration does not suppress structure formation and $f$ first flattens and then starts to grow again. This makes the value of the growth more similar to the $\Lambda$CDM model, thus explaining why, despite the unwanted feature of negative sound speed, the model represents a good fit to the data. In other words, since the instabilities arising at low redshift ($z\approx 0.33$), they did not have the time to considerably affect the evolution of perturbations. Finally, the time evolution of the growth index $\gamma$ (Fig.~\ref{ris:image03}) for models where the adiabatic sound speed and the DE EoS are approximately constant, is approximately constant as well, justifying our simplifying assumption. In general, $c_{\rm s,a}^2$ and $\omega_{\rm de}$ can be well approximated to a constant for the $\Lambda$CDM and $\omega$CDM models and for the logotropic models I--III, but this is not the case for the CPL parametrization and for the model IV introduced in this work. It is interesting to notice that for models I and III, $\gamma$ evolves like the corresponding adiabatic sound speeds. This is easy to understand as this is the only relevant quantity evolving in our analytical approximation. For model IV our approximation is no longer valid since both $\omega_{\rm de}$ and $\gamma$ evolve rather strongly with time. The effects of the negative adiabatic sound speed emerge in the behavior of the growth index: since its value reduces and it is about 30\% lower than for the $\Lambda$CDM model, the growth would be enhanced, despite the recent accelerated expansion.

Analyzing both panels in Fig.~\ref{ris:image03} more quantitatively, we see that our analytical prediction reproduces the numerical solution at the percent level, with the exception, as already discussed, of model IV. For this model, only its qualitative behavior is reproduced.

\section{Final remarks}\label{sezione6}
In this paper, we investigated four thermodynamic models that fuel current acceleration through a single fluid unifying DE with DM. The four models under examination are part of two typologies: Anton-Schmidt and logotropic fluids, intertwined among them by fixing the Gr\"uneisen index to precise values. These scenarios have recently been introduced as alternatives to the standard cosmological model with the purpose of extending modified Chaplygin gas models. In particular, we investigated the Anton-Schmidt paradigms for two relevant cases of the Gr\"uneisen index. Afterwards, we studied the logotropic framework, corresponding to $n=0$, and then we proposed a novel approach, namely model IV, in which DM and baryons are disentangled.  

We first discussed the basic features of each model and then evaluated the general picture of small perturbations, computing the growth $f=\mathrm{d}\ln{\delta}/\mathrm{d}\ln{a}$ (with $\delta$ the linearly perturbed density) and growth index $\gamma$ for each model. We portrayed and confronted the functional behaviors of the adiabatic sound speed, total and DE EoS and Hubble parameters for each model. 

Further, we compared our models with data at intermediate redshifts, through a MCMC procedure based on the Metropolis algorithm. To do so, we took into account SNeIa, OHD and linear growth observations for the large scale structures. We fixed tight constraints over the thermodynamic free parameters of our four models and we inferred the statistical significance of them through the AIC and BIC statistical criteria. 
We showed that the logotropic models behave better than Anton-Schmidt ones and the case $n=-1$ is fully ruled out.
Additionally, we demonstrated that the logotropic models are less stable than the standard cosmological paradigm, stressing that there is no statistical advantage to handle a logotropic and/or Anton-Schmidt universe, if compared with the $\Lambda$CDM, $\omega$CDM and CPL scenarios. In support of this, we outlined where perturbations become unstable due to negative sound speed for all the underlying models. 

In the future, we will investigate a possible time dependence of the Gr\"uneisen index $n$ for Anton-Schmidt models, as predicted by solid state physics, and check if it would enable the sound speed to be positive definite at all times. We will also attempt unifying DE with inflationary epochs through a single description based on Anton-Schmidt and/or logotropic models.

\begin{acknowledgments}
K. B., O. L. and M. M. acknowledge the Ministry of Education and Science of the Republic of Kazakhstan, Grant No. IRN AP08052311. MM is supported by INFN as part of the MoonLIGHT-2 experiment in the framework of the research activities of CSN2. F. P. acknowledges the support from the Grant No. ASI n.2018-23-HH.0.
\end{acknowledgments}

\bibliography{references}

\end{document}